# A Pair Production Telescope for Medium-Energy Gamma-Ray Polarimetry

Stanley D. Hunter[1a], Peter F. Bloser[b], Gerardo O. Depaola[c], Michael P. Dion[d], Georgia A. DeNolfo[a], Andrei Hanu[a], Marcos Iparraguirre[c], Jason Legere[b], Francesco Longo[e], Mark L. McConnell[b], Suzanne F. Nowicki[a,f], James M. Ryan[b], Seunghee Son[a,f], and Floyd W. Stecker[a]

[a] NASA/Goddard Space Flight Center, Greenbelt Road, Greenbelt, MD 20771

[b] Space Science Center, University of New Hampshire, Durham, NH 03824

[c] Facultad de Matemática, Astronomía y Fisica, Universidad de Córdoba, Córdoba 5008 Argentina

[d] Pacific Northwest National Laboratory, Richland, WA 99352

[e] Dipartimento di fisica, Università Degli Studi de Trieste, Treste Italy

[f] Department of Physics, University of Maryland Baltimore County, Baltimore, MD 21250

## ABSTRACT

We describe the science motivation and development of a pair production telescope for medium-energy (~5-200 MeV) gamma-ray polarimetry. Our instrument concept, the Advanced Energetic Pair Telescope (AdEPT), takes advantage of the Three-Dimensional Track Imager, a low-density gaseous time projection chamber, to achieve angular resolution within a factor of two of the pair production kinematics limit (~0.6° at 70 MeV), continuum sensitivity comparable with the Fermi-LAT front detector ($<3\times10^{-6}$ MeV cm$^{-2}$ s$^{-1}$ at 70 MeV), and minimum detectable polarization less than 10% for a 10 mCrab source in $10^{6}$ seconds.

---

[1]Corresponding author at: NASA/GSFC, Code 661, Greenbelt, MD 20771, USA. Tel: +1 301 286-7280; fax +1 301 286-0677, *E-mail address*: stanley.d.hunter@nasa.gov phone;

## 1. Introduction

Since the launch of AGILE [1] and Fermi/LAT [2], the scientific progress in high-energy ($E_\gamma \gtrsim 200$ MeV) gamma-ray science has been extensive. Both of these telescopes cover a broad energy range from ~20 MeV to >10 GeV. However, neither instrument is optimized for observations below ~200 MeV or for polarization sensitivity. Ground-based air Cherenkov telescopes have been used to observe both galactic sources such as supernova remnants and extragalactic sources of very high energy (TeV) gamma-rays such as active galactic nuclei (AGN) [3]. They have provided important astrophysical information, but they also lack the capability to detect polarization. The Fermi and AGILE space-based telescopes, operating in the GeV energy range, are expected to continue to make significant progress for the next several years. However, there remains a significant gap in our knowledge of astronomy in the medium-energy (~0.1–200 MeV) regime between the X-ray and high-energy gamma-ray energy ranges.

The next major step in gamma-ray astrophysics, recognized as early as the SAS-2 era [4], should be a medium-energy gamma-ray pair production telescope to fill this gap and provide answers to many important astrophysical questions. In the following, we describe the science motivation for this mission and the design of the Advanced Energetic Pair Telescope (AdEPT) a pair production telescope for medium-energy, ~5 to ~200 MeV, gamma-ray polarimetry.

## 2. Science Motivation

The AdEPT pair production telescope for the detection of medium energy (~5-200 MeV) gamma-rays with high angular resolution and polarimetry capabilities will open a new window in observational astronomy and astrophysics. Such an instrument can help provide answers to important questions in both astronomy and physics. For example, it can shed light on the origin and acceleration of cosmic rays, the nature of the cosmic-ray acceleration of electrons in the Crab nebula to energies in excess of $10^{15}$ electron volts [5] and how pulsars, with high magnetic fields and expected high gamma-ray polarization, achieve such high efficiency for particle acceleration. Gamma-ray polarization can distinguish between emission processes such as synchrotron radiation and other gamma-ray production mechanisms, however, the angular resolution with which the geometry of the gamma-ray emission regions are probed by polarization measurements is limited by the instrument angular resolution. It has long been expected that other astronomical sources such as "blazars" (a class of active galactic nuclei) should produce polarized gamma-radiation owing to the highly structured magnetic fields in their emission regions [6][7][8][9]. It is also known that gamma-ray bursts (GRBs) emit hard X-radiation whose polarization has been detected by space borne instruments, e.g. RHESSI [10], INTEGRAL [11], and GAP[12]. Such polarization should extend into the gamma-ray range, given the same basic emission processes. Observations at higher energies will investigate an underexplored energy range and provide new understanding of emission mechanisms with high polarization sensitivity.

Medium energy polarization measurements with AdEPT can also explore fundamental questions in theoretical physics. There is an apparent incompatibility between relativity and quantum mechanics at the Planck scale of $1.6\times10^{-35}$ meters. Effective field theory models developed to determine possible quantum gravity effects at observable energies, have led to the prediction of possible "vacuum birefringence", a process in which photons of different polarizations travel at slightly different velocities from an astronomical source. Such a process, if it exists at a significant enough scale, can destroy the inherent polarization of a source from which such polarization would be seen in its absence. Thus, the detection of polarization from a distant source such as a gamma-ray burst can constrain the possible existence of violations of relativity [13][14]. The birefringence effect is sensitive to the square of the photon energy. To date, the INTEGRAL/IBIS observations of the Crab pulsar and nebula at 200-800 keV [15] are the highest energy photon polarization measurements that have been made. An instrument capable of detecting polarization of medium energy gamma-rays can provide a much more sensitive probe of such relativity violations.

The AdEPT pair production telescope also has significant advantages over previous attempts to measure the medium-energy diffuse extragalactic gamma-ray background. Possible contributing components [16] include non-thermal tails from Seyfert galaxies, red-shifted lines from Type Ia and Type II supernovae, and unknown extragalactic sources. Measurements by both the Apollo21 [17] and COMPTEL [18] instruments were plagued by intrinsic detector and spacecraft background problems owing to the buildup of long-lived radioisotopes created by cosmic-ray interactions. The subtraction of such poorly determined backgrounds led to uncertainties in the extragalactic background determination and significantly different results reported by the two different instruments. A free-flying argon gas AdEPT instrument is expected to have low intrinsic background similar to EGRET and Fermi/LAT and therefore yield a more reliable determination of the extragalactic gamma-ray background in the medium-energy range.

The 5-plus fold improvement in angular resolution of AdEPT below ~200 MeV compared with Fermi/LAT, see Figure 1 will enable the numerous soft gamma-ray sources in the galactic plane to be better resolved improving the determination of the medium-energy Galactic diffuse emission and to spatially resolve variation between electron dominated and hadron dominated processes in the 70-200 MeV range.

## 3. Obtainable Goals for Exploring the Medium Energy Gamma-ray Universe

Significant improvement in sensitivity for pair telescopes can only be achieved through a dramatic improvement in the angular resolution, especially at lower energies. The ultimate angular resolution of any nuclear pair-production telescope is limited by the unobserved recoil momentum of the nucleus. The nuclear recoil momentum calculated by Jost, Luttinger & Slotnick [19] for photon energy $E_\gamma$ has a broad distribution extending from $2\mathrm{m_e^2}/E_\gamma$ to $E_\gamma$, where $\mathrm{m_e}$ is the electron rest mass, and the nuclear momentum is nearly orthogonal to the gamma-ray momentum. On the assumption that the recoil momentum is transverse to the photon direction [20], an upper limit to the kinematic limit can be defined as $q_{68}/E_\gamma$, where $q_{68}$ is the momentum above which 68% of the distribution lies. This assumption is not valid at energies below ~25 MeV where the momentum distribution is wider and the recoil angle is more acute. The kinematic limit and twice the limit are shown in Figure 1 as the solid and dotted magenta lines, respectively. In the case of triplet production, i.e. pair production on the atomic electrons, the recoil momentum is, in most cases, observable [21] and the angular resolution is limited by the energy and spatial resolution of the electron track imager. Further discussion of triplet detection with AdEPT including effective area (enhanced for low-Z materials), angular resolution, and polarization asymmetry factor is beyond the scope of this paper and will be addressed in a future paper.

The performance goals of a telescope to address the questions outlined above plus a wide variety of other interesting topics including solar flares, diffuse emission, etc. are summarized in Table 1.

| Table 1 – AdEPT Instrument Performance Goals |
|---|

| | |
|---|---|
| Energy range | ~5 to ~200 MeV |
| Energy resolution | ~30% ΔE/E (FWHM) at ~70 MeV |
| Angular resolution | < 1° at 70 MeV |
| Continuum sensitivity×$E^2$ | ~$3\times10^{-5}$ MeV $cm^{-2}$ $s^{-1}$ |
| Minimum detectable polarization | ~10% for 10 mCrab in $10^6$ s (asymmetry factor, λ = 0.15) |

In the following, we show that a medium energy gamma-ray telescope meeting these angular resolution and polarization sensitivity requirements can be developed and describe the design of the Advanced Energetic Pair Telescope (AdEPT)[2] the telescope presently being developed using the Three-Dimensional Track Imager (3-DTI) technology developed at Goddard Space Flight Center (GSFC). In the following sections we describe the advantages of the 3-DTI gas detector technology and the predicted performance of AdEPT.

## 4. Advantages of a Gas Detector for a Pair Production Telescope

The design of all pair production space telescopes to date, SAS-2 [22], COS-B [23], EGRET [24], AGILE [1], and Fermi-LAT [2] have utilized an electron tracking hodoscope consisting of a stack of electron tracking detectors interleaved with metal foils, each typically ~20 milliradiation lengths (mRL) thick, positioned above a calorimeter. SAS-2, COS-B and EGRET utilized two-dimensional gas spark chambers whereas AGILE and Fermi/LAT have taken advantage of silicon-strip detectors (SSD). The multiple layers of high-Z metal foils (totaling about 500 mRL) provide substantial material for high interaction probability and large effective area, however, they also contribute to multiple Coulomb scattering (MCS) which degrades the accuracy with which the electron and positron directions emanating from the pair vertex can be determined. Kryshkin, Sterligov, & Usov [25] determined that these directions for high energy (900 MeV) gamma rays, which form the basis of gamma-ray direction and polarization determination, are dominated by MCS after traversing about 10 mRL of material. The maximum material thickness would be even less for lower energy gamma rays.

In low-Z material, $Z \lesssim 30$, gamma rays with energy below ~10 MeV, are more likely to interact via Compton scattering than pair production, however, the intrinsic modulation factor of polarized gamma rays interacting via pair production is higher above ~2 MeV, compared to Compton scattering and photo-electric absorption [26]. Thus, we are motivated to reduce the effective minimum energy of a pair telescope towards the threshold energy, to take advantage of the higher modulation factor. This requires that the direction of the electron and positron emanating from the pair vertex, which forms the basis of the gamma-ray direction and polarization determination, be measured in less than ~10 mRL of material after which their directions are dominated by multiple Coulomb scattering (MCS) [25].

[2] A list of less common acronyms is given at the end this paper.

In the remainder of this section we give a detailed calculation that corroborates the conclusion that a low density, less than ~5 mg/cm$^3$, track imager is required to achieve the AdEPT performance.

*4.1 Electron track measurement constraint*

Achieving high angular resolution and the lowest possible minimum detectable polarization (MDP) requires a new approach to reduce the density of the conversion and scattering material per track measurement interval in the hodoscope. The density per measurement interval (measurement density) of a hodoscope with interleaved foils can be reduced by decreasing the thickness of the conversion material or increasing the separation between the measurement layers.

The concept of reducing the thickness of the converter material in a gamma-ray telescope to improve the medium-energy sensitivity was recognized by Kniffen, et al. [27]. They achieved nearly an order of magnitude increase in sensitivity at 20 MeV by replacing the lead conversion foils in a gas spark-chamber telescope, used previously for high-energy gamma-ray observation [28], with aluminum foils. More recently, several pair telescopes have been proposed without conversion foils, i.e. the conversion material is the SSDs of the track imager. Proposed applications of this concept are MEGA [29], TIGRE [30], and a GLAST/LAT modification [31].

The AdEPT gamma-ray telescope concept (Table 3) takes advantage of a gaseous medium to provide a homogenous tracking detector to achieve nearly continuous measurements of the electron and positron tracks from pair production. The optimal fit formulas derived by Innes [32] for estimating the tracing parameter error matrix in the case of a homogenous detector with many layers, can be used to estimate the AdEPT angular resolution and place an upper limit on the gamma-ray convertor density.

Innes describes the projection of a nearly straight track onto the plane perpendicular to the magnetic field as $y = a + bx + cx^2/2$ where $a$ is the position at the beginning of the track, $b$ is the slope, $c$ is the curvature, and $x$ is the distance along the track. Innes defines the information density of a detector with total length $L$, $N$ equally spaced measurement layers, and RMS measurement error at each layer $\sigma_{\mathrm{m}}$ as $\iota = (N + 5)/\sigma_{\mathrm{m}}^2 L$. The information density of the EGRET [33], MEGA [29], Fermi-LAT front section (hereafter Fermi/LAT) [34], and AGILE [35] telescopes along with the AdEPT concept (Table 3) are listed in Table 2. We have restricted the AdEPT detector length, corresponding to selection of the initial portion of the electron track, and taken an upper limit to the measurement spacing, see §5.2.

Table 2 – Gamma-ray Telescope Information Density

| Telescope | $N$ | $L$ (mm) | $l$=$L$/$N$ (mm) | Pitch (mm) $\sigma_{\mathrm{m}} = P/\sqrt{12}$ | $\iota$ (mm$^{-3}$) |
|---|---|---|---|---|---|
| EGRET | 28 | 450 | 16.1 | 0.810 | 1.33 |
| MEGA | 32 | 320 | 10.0 | 0.470 | 6.28 |
| Fermi/LAT | 12 | 416 | 34.7 | 0.228 | 9.42 |
| AGILE | 12 | 228 | 19.0 | 0.242 | 17.38 |
| AdEPT | 300 | 300 | 1.0 | 0.400 | 76.25 |

Innes further defines the characteristic scattering length $\ell = 1/\sqrt[4]{\iota s}$ in terms of the information density and $s$ the mean square projected scattering angle per unit length. Omitting the small logarithmic term in the Gaussian approximation for the central 98% of the projected scattering angular distribution [36][37], we have $s \approx \left({}^{13.6\ \mathrm{MeV}}/_{\beta cp}\right)^2 \cdot {}^{1}/_{X_0'}$, where $X_0' = X_0/\rho$ is the RL/density in units of mm. Innes derives formulas (Eq. 11 of [32]) for the optimal fit variance matrix elements in the *continuous detector limit* ($l \to 0$, with $\iota$ and $s$ constant, for large $N$) in the limiting case, $u = L/\ell > 7$ where the tracking error is dominated by MCS, and $u < 7$, where the error is determined by the tracer spatial resolution. For the AdEPT concept described in Tables 2 & 3, $u$ is well approximated by $u \approx 60.9 \cdot \sqrt{{}^{10\ \mathrm{MeV}}/_{E_e}}$ for electron energies above ~0.5 MeV. Over the medium-energy range, $u \gg 7$ and the AdEPT angular resolution is dominated by MCS and we do not consider the correction to the Innes equations for $u < 7$ in the absence of a magnetic field noted by Bernard [38]. For nearly straight tracks in the absence of magnetic field curvature, the variance of the optimum fit track slope is given by the $V_{bb}$ matrix element

$$\sigma_{\theta t}^{2} = V_{bb} \approx {}^{1}\!/_{\iota\ell^{3}} \left({}^{2}\!/_{u} + \sqrt{2}\right) \xrightarrow{u \gg 1} {}^{\sqrt{2}}\!/_{\iota\ell^{3}}\,. \quad (1)$$

We note that this formula is not applicable to MEGA, EGRET, Fermi/LAT, and AGILE because of their small number of measurement layers and discrete convertor foils.

The gamma-ray direction, in the small-angle approximation, is reconstructed by combining the measured directions of the electron and positron weighted by their energies (see §3.11 of [20]). In the multiple scattering dominated regime, the conversion of track resolution to photon angular resolution is $\sigma_{\theta_\gamma} = \sigma_{\theta_t}\sqrt{\sqrt{r} + \sqrt{1-r}}$ where $r$ and $1 - r$ are the energy fractions carried away by the electron and positron. In the case of equal energy partition ($r = 0.5$) we have $\sigma_{\theta_\gamma} \approx 1.2\ \sigma_{\theta_t}$, which becomes $\sigma_{\theta_\gamma} = \sigma_{\theta_t}$, for asymmetric energy partition, $r \approx 0$.

The point spread function of a gamma-ray telescope is often expressed in terms of the 68% containment angle $\theta_{68}$, which is related to the standard deviation of a Gaussian distribution by

$$\theta_{68} = \sqrt{-2 \cdot \ln(1 - 0.68)}\, \sigma_{\theta\gamma} \approx 1.51\ \sigma_{\theta\gamma}. \quad (2)$$

The expected angular resolution of the AdEPT telescope is the quadrature sum of the kinematic limit and $\theta_{68}$ and is shown as the solid black line in Figure 1. Up to some energy the angular resolution is dominated by the kinematic limit. We denote the gamma-ray energy at which the angular resolution is twice the kinematic limit, indicated by the vertical black dotted line in Figure 1, as $E_{\gamma,\mathrm{KL}} \approx 150$ MeV. The corresponding angular resolution, $\theta_{68,\mathrm{KL}} \approx 6$ mrad, indicated by the horizontal black dotted line determines the electron-positron direction error $\sigma_{\theta t,\mathrm{KL}} \approx 3.36$ mrad and, in the limit $u \gg 1$, Eq. (1) can be used to estimate the scattering angle, *s*, and hence the RL of the electron-positron tracking medium.

$$s = \sqrt[3]{\sigma_{\theta t}^{8}\ \iota/4} \approx \left(\frac{13.6}{\beta cp}\right)^{2} \frac{1}{X_0'} \quad (3)$$

Using the $\theta_{68}$ value from Figure 1 and assuming $E_e = E_\gamma/2 - \mathrm{m}_e$ the value of $X_0'$ increases from ~$0.15 \times 10^4$ cm at 5 MeV to ~$1.75 \times 10^4$ cm at ~800 MeV. On the assumptions, $\mathrm{A} \approx 2.1\ \mathrm{Z}$ for low-Z materials ($\mathrm{Z} < 20$), $\mathrm{A} \approx 2.5\ \mathrm{Z} - 8$ for higher Z materials ($20 < Z < 54$), and using Dahl's approximation [37] for the RL of a material, $\mathrm{X}_0 = \frac{716.4\ \mathrm{A}}{\mathrm{Z}(\mathrm{Z}+1)\ln(287/\sqrt{\mathrm{Z}})}\ \mathrm{g/cm^2}$, the density of the tacking material must lie between ~0.5 and ~5 mg/cm$^3$. With the exception of modern aerogel materials [39] this low density can only be realized with a gaseous media.

*4.2 Polarization constraint*

The use of the azimuthal orientation of the electron positron plane to determine the gamma ray polarization has been discussed since the 1950s [40][41] and imposes another constraint on the track imager. In the case of small nuclear recoil, i.e. the electron, positron, and incident photon are nearly coplanar, the azimuthal dependence of the cross section can be written in the form [42]

$$\sigma(\phi) = \frac{\sigma_0}{2\pi}[1 + P\lambda\cos^2\phi] \tag{4}$$

where $\sigma_0$ is the total cross section, $\phi$ is the angle between the electron-positron pair and the photon's electric field vector, $P$ is the fractional polarization, and $\lambda$ is the inherent azimuthal asymmetry factor. Kel'ner, Kotov, and Logunov [26] calculated the azimuthal asymmetry of the secondary emission for the photoelectric, Compton, and pair production processes. They found that the asymmetry factor λ is higher for photoelectric and pair production processes below and above ~2 MeV respectively compared to the Compton process and they showed that the polarization modulation factor and hence the polarization sensitivity, decreases exponentially with the thickness of material traversed by the pair electrons after only a few mRL [26]. These calculations assume that the electron and positron energies are greater than 1.5 MeV and that the angle between them and the photon is less than 40°. These calculations support the conclusion of Mattox [43][44] that the thickness of the conversion foils in previous telescopes (typically 20 mRL) precludes any polarization sensitivity for these instruments. Buehler et al. [45] estimated that Fermi LAT might have marginal polarization sensitivity by selecting 50-200 MeV photons that convert in the silicon rather than the 30 mRL thick tungsten foils. Their analysis, omitting background and trial factor considerations, concluded that 20% polarization from Vela could be detected at 3σ using 20 months of data. The 8 mRL thickness of the two silicon-strip detectors reduces the polarization sensitivity.

*4.3 Gaseous track imager*

Significant advances in medium-energy gamma-ray pair production telescopes can only be realized if the density of the material in the track imager is drastically reduced. A low-density, homogenous detector that provides high spatial resolution tracking and substantially minimizes the effects of Coulomb scattering is required. These technical challenges were met historically with a whole genre of detectors based on gas physics including cloud chambers, bubble chambers, and gas-wire detectors (spark and drift chambers, etc., see e.g. [46]). The use of a gaseous medium time projection chamber (TPC) as both the conversion and detection medium for a gamma ray pair telescope was first suggested by Hartman [47] and further explored by Bloser, et al. [48][49]; Hunter, et al. [50][51]; Ueno, et al. [52], and Bernard [20]. In the following section we describe our Three-Dimensional Track Imager (3-DTI), a gaseous time projection chamber technology, its application to the design of AdEPT, and the expected performance of the AdEPT pair polarimeter instrument.

## 5. Three-Dimensional Track Imager (3-DTI)

The 3-DTI detector, shown schematically in Figure 2, combines a gas Time Projection Chamber (TPC) [53][54] and a 2-D readout to provide a low density gamma-ray conversion medium with high-resolution, 3-D charged particle tracking obtained by digitizing the 2-D readout signals. The 3-DTI also takes advantage of Negative-Ion drift [55] to reduce diffusion to the thermal limit without an applied magnetic field allowing the TPC drift distance to be much larger than would be possible with free electron diffusion.

The TPC volume, which defines the 3-DTI active volume, is bounded by a drift electrode on the top, a linear potential gradient field-shaping cage of wires, and 2-D readout plane on the bottom. A charged particle traversing the gas medium loses energy by ionization. The ionization electron density is proportional to the $\mathrm{d}E/\mathrm{d}x$ energy loss of the particle along its track. The drift field, the electric field in the TPC active volume (~1 kV/cm), causes the ionization to drift at a uniform velocity onto the 2-D readout plane. The relative 3-D location of the ionization charge is determined from the 2-D readout and time of arrival.

### *5.1 Two-Dimensional Readout*

The 2-D readout consists of a 2-D Micro-Well Detector (MWD) [56][57][58][59][60] with pre-amplification provided by a Gas Electron Multiplier (GEM) [61]. Two-stage amplification of the ionization charge was required to detect single ionization electrons using the negative ion drift technique (see §§5.2, 5.3). The MWD consists of two orthogonal layers of electrodes separated by an insulating substrate, see lower right inset in Figure 2. The cathode and anode electrode strips are etched onto the top and bottom layers respectively of a copper clad insulator using flex circuit board technology. Holes etched in the top (cathode) electrodes are concentric with blind vias micro-machined in the insulator that expose the anode electrode and form the micro-wells. Our MWD design has 200 $\mu$m diameter × 200 $\mu$m deep wells on 400 $\mu$m × 400 $\mu$m center-to-center pitch. Charge entering a micro-well is accelerated by the strong electric field (~40 kV/cm) in the wells and can produce a Townsend electron avalanche (e.g. [62]) proportional to the primary ionization charge. This amplification, or "gas gain", is exponentially dependent on the electric field in the micro-well. The avalanche electrons are collected on the anode and the motion of the avalanche charge induces an equal but opposite image charge on the cathode. The anode and cathode signals provide the 2-D, (X-Y), spatial location of the primary ionization. Sampling of the avalanche charge signal at a fixed frequency allows the third dimension Z (height) to be calculated from the uniform drift velocity of the ionization charge through the gas volume.

*5.2 Negative Ion Diffusion*

Diffusion of the ionization electrons drifting through the gas places an upper limit on the useful height of the TPC. If the maximum allowable diffusion is chosen to be twice the TPC readout pitch, then the maximum drift distance can be expressed in terms of the diffusion coefficient: $Z_{\max} = (2\ \mathrm{Pitch}/\sigma_0)^2$. For example, Puiz [63] measured the electron drift velocity and diffusion in an Ar+$CO_2$ (80%/20%) mixture at 1 atm. The drift velocity of free electrons $V_d$ increases quasi-linearly with the drift field, $E$, with reduced mobility $\mu = V_d/E \approx 4.2 \times 10^3\ \mathrm{cm^2 atm/Vs}$ (Fig. 17 of [63]). The electron diffusion coefficient, $\sigma_0$, shown as the red line in Figure 3, exhibits thermal behavior decreasing as $1/\sqrt{2kT/eE}$, blue line, up to ~100 V/cm. For higher drift fields, the electron drift velocity is significantly higher than the thermal velocity of the gas and $\sigma_0$ tends to increase with $E$ reaching a plateau at high fields $\gtrsim 800\ \mathrm{V/cm\ atm}$. The minimum diffusion value of $\sigma_0 \approx 180\ \mu\mathrm{m}/\sqrt{\mathrm{cm}}$, is reached at ~300 V/cm, corresponding to a drift velocity $V_d = \mu E \approx 1.2 \times 10^4\ \mathrm{mm/ms}$. The maximum drift distance in Ar+$CO_2$ is then ~20 cm for a detector pitch of 400 $\mu$m.

Thermal diffusion can be maintained at higher fields by adding an electronegative component to the gas that captures the primary ionization electrons, forming negative ions, which then drift in thermal equilibrium with the gas. Carbon disulphide ($CS_2$), with a vapor pressure of ~300 torr at 300 K and moderate electron affinity [64], has successfully been used as a negative ion molecule in the Negative Ion-Time Projection Chamber (NI-TPC) [55]. We have measured the negative ion diffusion coefficient in $CH_4$+$CS_2$ [65] and in Xe at 100, 200, 300 torr plus 40 torr $CS_2$. The Xe results are shown in Figure 3. For both gas mixtures, the negative ion diffusion coefficient is reduced to the thermal limit and is a function only of the drift field and gas temperature, and independent of the gas mixture. For negative ions, the diffusion coefficient decreases with $E$ becoming less than $\sim 80\ \mu\mathrm{m}/\sqrt{\mathrm{cm}}$ above ~800 V/cm at 25°C. In this case, the maximum drift distance for a detector pitch of 400 μm is greater than ~100 cm. We note that the diffusion can be further reduced by operating the TPC at lower temperature, e.g. 0°C or lower.

The negative-ion drift technique allows for large TPC active volumes that can be read out with one readout layer, however, the drift velocity is substantially lower than for free electrons and, for constant voltage, the gas gain with $CS_2$ added is reduced about 100-fold. The drift velocity of the $CS_2^-$ ions in Xe is ~10 mm/ms at 800 V/cm; about 3 orders of magnitude slower than that of free electrons in Ar+$CO_2$ [63]. Reduced drift velocity is an advantage in the digitization rate corresponding to a z-coordinate resolution is reduced and the sampling rate of the digitizers can also be reduced which results in lower instrument power, see §6. $CS_2$ also provides strong UV quenching, which reduces breakdown brought on by the electron avalanche and ensures stable operation. The threshold for electron dissociation of $CS_2$ is 9.337±0.06 eV [66], thus, dissociation represents a negligible effect and gas degradation should be minimal ensuring a long instrument life-time.

Since thermal diffusion is independent of the gas mixture, we choose Ar+$CS_2$ rather than Xe+$CS_2$ for 3-DTI because of the higher drift velocity, reduced Coulomb scattering, and higher relative triplet production. Our measured mobility in Ar+$CS_2$ at 660 torr and 1200-1500 V/cm is 16-20 mm/ms, consistent with the results of Ohnuki et al. [67].

*5.3 Single Ionization Electron Detection*

Detection of the ionization electrons along the tracks of the electron and positron pair is a requirement for gamma-ray imaging. Generation of a Townsend avalanche in the MWD requires detachment of the ionization electrons from the negative ions. This occurs in a strong electric field of the micro-well by collision of the negative ion with the gas molecules [68]. The free electrons are then accelerated in the micro-well producing the avalanche. The start of the avalanche from a detached electrons, however, occurs lower in the micro-well (closer to the anode) compared with a free electron avalanche resulting in lower gain for given MWD voltages, shown in Figure 4. The maximum electric field, and hence maximum avalanche gain, is determined by the maximum stable operating voltage of the micro-well detector corresponding with Raether's criterion in micro-pattern gas detectors. This was demonstrated by Ivaniouchenkov et al. [69] and Bressan et al. [70]. The micro-well voltage cannot be increased sufficiently to overcome the reduction in gain caused by the negative ion collision effect. We have overcome the reduction in gain by adding a Gas Electron Multiplier (GEM) [61] pre-amplification stage to our micro-well detector. The gain of our MWD with and without the GEM pre-amplifier measured in P-10 (90% Ar + 10% $CH_4$) and Ar+$CS_2$ (560 torr+40 torr) at a total pressure of 600 torr is shown in Figure 4. Gains in excess of $10^4$ are readily achievable with the MWD+GEM combination in Ar+$CS_2$ providing single ionization electron detection. The X-Z projection of the electron-positron tracks resulting from the pair interaction of a 6.129 MeV gamma ray is shown in Figure 5 obtained with a small 5×5×9 $cm^3$ 3-DTI prototype with MWD+GEM readout. These highly structured tracks show pulse amplitude variation proportional to the $\mathrm{d}E/\mathrm{d}x$ energy loss of the electrons along their paths, multiple Coulomb scattering, and the formation of the Bragg peak of the stopping lower energy particle. X-Z and Y-Z projections of typical electron track from $^{90}$Sr are shown in Figure 6. The $\theta_{68}$ value derived from a very preliminary angular resolution measurement, based on only a few 6.129 MeV interactions, was ~18 deg. This measurement is ~2.5 times greater that the kinematic limit. Agreement is quite good given that the electron track lengths were short and no correction was made for near-field parallax.

*5.4 3-DTI Prototype Development*

The development of the 3-DTI has been done in stages, our 10×10×15 $cm^3$ and 30×30×15 $cm^3$ versions are shown in Figure 7. A 30×30×7 $cm^3$ 3-DTI detector was used for an Office of Naval Research funded demonstration of neutron imaging [71] in an over-water environment. The 2-D readout for the 3-DTI detector used for neutron imaging did not require the two-stage GEM pre-amplifier because of the much higher specific ionization of protons compared to minimum ionizing electrons. We are in the process of expanding our mechanical support technique for the MWD+GEM to 10×10 $cm^2$ and 30×30 $cm^2$ MWDs. These larger prototypes will be used to make much more detailed and accurate angular resolution measurements than described above. These measurements and comparison with the calculations in this paper will be reported on in a subsequent publication.

## 6. Design of the AdEPT Medium-Energy Pair Polarimeter

The design of the AdEPT pair polarimeter has matured along with the development of the 3-DTI detector technology. Our baseline concept for the AdEPT instrument and spacecraft is described in Table 3.

Table 3 - AdEPT Instrument and Spacecraft Concept

| | |
|---|---|
| Configuration: | 2 layers, 2×2×1 $m^3$ 3-DTI modules |
| $A_{geom}$: | 2×2 $m^2$ = 4×$10^4$ $cm^2$ |
| Depth: | 200 cm |
| 3-DTI Gas: | Ar (1100 torr) + $CS_2$ (40 torr) at 25 C |
| 3-DTI resolution: | 400 μm in x, y, and z |
| Pressure vessel: | Al, ~5 mm thick, ~300 cm diameter |
| Readout channels: | ~40,000 |
| Instrument power: | ~500 W |
| Instrument mass: | ~730 kg |
| Spacecraft: | zenith pointed, 3-axis stabilized |
| Orbit: | 28 deg, ~550 km altitude |

The total mass of the Ar+$CS_2$ gas is ~20 kg at 25°C with a corresponding RL of ~6.1×$10^3$ cm. The 3-DTI drift field will be ~1 kV/cm resulting in a negative drift velocity $v_D$ of ~18 mm/ms. The 3-D spatial resolution is determined by the MWD pitch and the sampling frequency of the analogue signals from the MWD. The 400 μm pitch of the MWD corresponds to a RMS resolution, $\sigma_{x,y}$ of $400\ \mu\text{m}/\sqrt{12}\ \approx 115\ \mu\text{m}$. Similar $z$ resolution of 400 μm is determined by the digitization rate which, with a five-fold over sampling to avoid aliasing, is determined by the negative ion drift velocity and the digitization rate is $5\ v_D/\sigma_{x,y} \approx 225$ kHz.

The slow negative ion drift velocity reduces the value of a charged particle anti-coincidence detector or calorimeter in the AdEPT design. The time required for the ionization (track information) associated with a cosmic-ray or electron/positron pair to drift to the readout layer, the read-out delay, is tens of milli-seconds. Thus, a temporal coincidence between the track information and fast scintillator pulses is impractical. An exception to the long read-out delay is for those tracks which traverse the readout plane. In this case the ionization charge closest to the readout layer is read out in the next digitization period, ~4 μs delay, and coincidence, in particular, with a calorimeter, may be practical.

In either case, an anti-coincidence signal cannot be used to discriminate cosmic rays from gamma-ray interactions since their ionization charge is in the gas and will be readout along with the gamma-ray track information. Thus, the readout delay results in a track "memory" in the TPC volume. In low-earth orbit, the integral SPENVIS[3] isotropic cosmic ray proton flux is ~56.6 $(m^2 \text{ sr s})^{-1}$ above ~6 GeV. The number of proton tracks crossing 1 $m^2$ face of the TPC, with acceptance of $1\pi$ $m^2$ sr, in a time corresponding to the maximum drift time, 1000 mm/$V_d$, is ~10 tracks/face or ~60 tracks/$m^3$. The Gamma-ray interactions must be identified and discriminated from these tracks, using image recognition techniques. In separate work [72], a multi-core processor has been demonstrated and software is being developed to process the Giga-bit per second raw data from a 1 $m^3$ TPC and separate the gamma-ray tracks from the CR tracks. Initial processing of simulated AdEPT data indicates that this separation, due to the high spatial information provided by the 3-DTI, is nearly lossless and result in little loss of effective area.

At this point in the development of the AdEPT concept we omit the anti-coincidence because it is not effective and do not include a calorimeter. A calorimeter could be added later, at the expense of increased mass, instrument complexity, and reduced instrument solid angle, if further instrument optimization and mission studies warrant.

### *6.1 Effective area*

The performance of the AdEPT pair polarimeter has been calculated based on the concept parameters in Table 3, consideration of event reconstruction effects have not been included. The effective area of AdEPT is given by

$$A_{\text{eff}}(E_\gamma) = A_{\text{geom}} \cdot [1 - \exp(-\mu_{\text{pair}} \cdot \rho_{\text{gas}} \cdot \text{D})]\,. \tag{5}$$

Where $A_{\text{geom}}$ is the TPC geometric area, $\mu_{\text{pair}}(E_\gamma)$ is the pair interaction coefficient as a function of gamma-ray energy in $cm^2/g$, $\rho_{\mathrm{gas}}$ is the gas density in $g/cm^3$, and D is the depth of the TPC in cm. The effective area of AdEPT is plotted in Figure 8 as a function of gamma-ray energy. The EGRET [73] and Fermi/LAT front [74] effective areas are shown for comparison.

---

[3] Space Environment Information System https://www.spenvis.oma.be/

*6.2 Continuum sensitivity*

The continuum sensitivity corresponding to a significance level, $n_\sigma$, is calculated from the expression for effective source counts [75]

$$n_\sigma = {}^{S}\!/_{\sqrt{S+B}}\,. \quad (6)$$

Where $S$ and $B$ are, respectively, the number of source and background photons detected by an instrument with effective area $A_{\mathrm{eff}}$, in observation time $T_{\mathrm{obs}}$, and energy interval $\Delta E_\gamma$.

The easily recognized "Λ" signature of pair production results in detectors that are nearly free of instrumental background [47], thus the background counts are modeled using the all-sky average extragalactic gamma-ray emission spectrum derived from the EGRET data [76]

$$F_B(E_\gamma) = 7.32\times10^{-9}\left(E_\gamma/451\ \mathrm{MeV}\right)^{-2.10}\ \text{photons cm}^{-2}\ \text{s}^{-1}\ \text{sr}^{-1}\ \text{MeV}^{-1}.$$

We choose to use the flatter EGRET diffuse spectrum since it is consistent with the COMPTEL [77] and SAS-2 [78] results rather than the steeper Fermi spectrum [79], which if extrapolated down to the medium-energy region is inconsistent with the COMPTEL results.

The number of background counts is given by

$$B(E_\gamma) = F_B(E_\gamma)\cdot A_{\mathrm{eff}}(E_\gamma)\cdot T_{\mathrm{obs}}\cdot\Omega(E_\gamma)\cdot\Delta E_\gamma\ \text{photons}, \quad (7)$$

where

$$\Omega(E_\gamma) = 2\pi\left(1-\cos\theta_{68}(E_\gamma)\right)\ \mathrm{sr} \quad (8)$$

is the solid angle containing 68% of the photons from a point source. The number of source counts corresponding to detection significance is determined by solving Eq. (6) for $S(E_\gamma)$ and taking the positive root.

$$S(E_\gamma) = \frac{1}{2}\left(n_\sigma^2 + \sqrt{n_\sigma^4 + 4B(E_\gamma)n_\sigma^2}\right) \approx n_\sigma^2 + B(E_\gamma). \quad (9)$$

The corresponding differential continuum source flux or sensitivity is given by

$$F_S(E_\gamma) = S(E_\gamma)/(A_{\text{eff}} \cdot T_{\text{obs}} \cdot \Delta E_\gamma \cdot 0.68) \text{ photons cm}^{-2}\text{ s}^{-1}\text{ MeV}^{-1} \quad (10)$$

where the factor of 0.68 in the denominator corresponds to the use of $\theta_{68}$ in Eq. 8. The AdEPT $3\sigma$ continuum source sensitivity multiplied by $E_\gamma^2$ calculated on the assumptions of an observation time of $10^6$ s and $\Delta E_\gamma = E_\gamma$ is shown in Figure 9. The sensitivity for EGRET and Fermi-LAT front, calculated on the same assumptions are shown for comparison. The AdEPT sensitivity is better than Fermi up to ~200 MeV, as desired.

The AdEPT sensitivity calculated here must be considered as a lower limit since the effective area calculation does not include any corrections for interactions near the edge of the TPC or inefficiencies in the track recognition. Further, the sensitivity for sources near the Galactic center will be up to an order of magnitude higher, since the 30 to 100 MeV Galactic diffuse emission is about an order of magnitude higher than the extragalactic emission [80]. The assumption of low instrumental background may also be optimistic, since, without an anti-coincidence, neutral pions generated by cosmic-ray protons interacting with the pressure vessel are a potential source of background. This will be taken into account as part of the detailed instrument simulations.

*6.3 Minimum detectable polarization*

The minimum detectable polarization (MDP) for a given instrument can be written as [81] :

$$MDP(E_\gamma) = {}^{n_\sigma}\!/_{\lambda}\; {}^{\sqrt{S+B}}\!/_{S} \quad (10)$$

where S and B are the observed source and background counts and the asymmetry factor, is defined in Eq. 4. The asymmetry factor for co-planner events has been calculated over the entire energy range allowing for screening of the nucleus by Kel'ner et al. [26] and above 10 MeV using Monte Carlo integration by Depaola and Kozemeh [82]. The Kel'ner value rises rapidly from zero at 1 MeV to a maximum of ~0.46 at ~2 MeV and then decreases to a high energy asymptotic value of ~0.4. The Depaola and Kozemeh value is ~0.12 at 10 MeV and rises to an asymptotic value of ~0.2. The difference in asymptotic values may be due to different assumptions made in the calculations. Depaola and Kozemeh also find that the asymmetry factor also changes sign for events with small deviations from co-planarity, thus to obtain a more accurate value of the asymmetry factor it will be necessary to include the instrument angular resolution in the integration. For this work we adopt a conservative fixed value for the asymmetry factor $\lambda = 0.15$ for $E_\gamma > 5$ MeV to evaluate the MDP for a source with spectrum and intensity equal to the Crab nebula [83] (defined as a "1 Crab" source). Assuming equal energy split between the electron and positron, the MDP for 1 Crab, 100 mCrab and 10 mCrab sources is shown in Figure 10.

*6.4 Energy resolution*

The high spatial resolution of the 3-DTI tracker allows multiple Coulomb scattering to be used to estimate the energy ($pv$) of electrons above ~1 MeV which will generally exit the 3-DTI gas volume. Specific ionization, $dE/dx$, and residual range can be used for lower energy and stopping electrons. The techniques developed to determine the energy of particles leaving tracks in photographic emulsions [84][85][86] have also been used with bubble chambers [87]. These techniques have been extended to include Kalman filtering [88] and been used to measure the through-going particle momentum in the ICARUS T600 TPC [89]. An estimate of the AdEPT energy resolution can be obtained by scaling the ICARUS muon momentum by the square root of the ratio of the detector RLs, i.e. $\sqrt{X_{0,\mathrm{L-Ar}}/X_{0,\mathrm{G-Ar}}} \approx 29.4$. This approach is valid because the spatial resolution per RL of AdEPT, $(1.2\times10^4$ cm/RL)/(0.04 cm/sample) = $3.0\times10^5$ samples/RL, is much higher than for ICARUS, (14 cm/RL)/(0.3 cm/sample) = 46.7 samples/RL. The scaled ICARUS simulations using the classical and Kalman filter methods, shown in Figure 11 with the scaled electron momentum, indicates that the expected AdEPT momentum resolution will be better than ~15% for electron momenta above ~10 MeV. The much higher resolution of AdEPT may result in improved low momentum resolution compared to ICARUS. Monte-Carlo simulations similar to those done for ICARUS are being done for the AdEPT instrument and the results will be presented in a future paper.

## 7. Summary

The AdEPT instrument concept based on the 3-DTI gas TPC technology (Table 3) will provide unique observations in the 5 to 200 MeV energy range. These observations with angular resolution within a factor of two of the pair production kinematic limit and minimum detectable polarization <2% for a 100 mCrab source up to ~150 MeV will address a wide range of the critical science goals. The calculated AdEPT performance is encouraging and the few instrument challenges are readily tractable. Detailed Geant4 simulations will be completed to confirm these calculations.

Milestones in the AdEPT development program include testing of $10\times10\times15\ cm^3$ and $30\times30\times30\ cm^3$ 3-DTI detectors. Currently proposed work will build a $50\times50\times100\ cm^3$ 3-DTI prototype of the AdEPT instrument module. Future work will include calibration of this AdEPT module prototype at the Duke University HIGS accelerator that offers 100% polarized gamma-rays from ~15 to 50 MeV [90], and a balloon flight in the 2018-20 time-frame. The goals of the accelerator calibration will be to determine the optimum electron energy determination algorithms, gamma-ray direction and energy, and the energy dependent polarization modulation factor, and to verify the angular and energy distributions for pair production near threshold simulated with Geant4. The balloon flight will confirm that gamma-rays can be identified in the presence of a high charge particle background.

We envision AdEPT, a future space mission, to be the next step in observational gamma-ray astrophysics that will open up a new window in medium-energy gamma-ray astrophysics with its unique capability to measure polarization. .

## ACRONYMES

| | |
|---|---|
| 3-DTI | Three Dimensional Track Imager |
| AdEPT | Advanced Energetic Pair Telescope |
| GEM | Gas Electron Multiplier |
| MCS | Multiple Coulomb Scattering |
| MDP | Minimum Detectable Polarization |
| mRL | milliradiation Length |
| MWD | Micro-Well Detector |
| NI-TPC | Negative Ion TPC |
| SSD | Silicon Strip Detector |
| TPC | Time Projection Chamber |

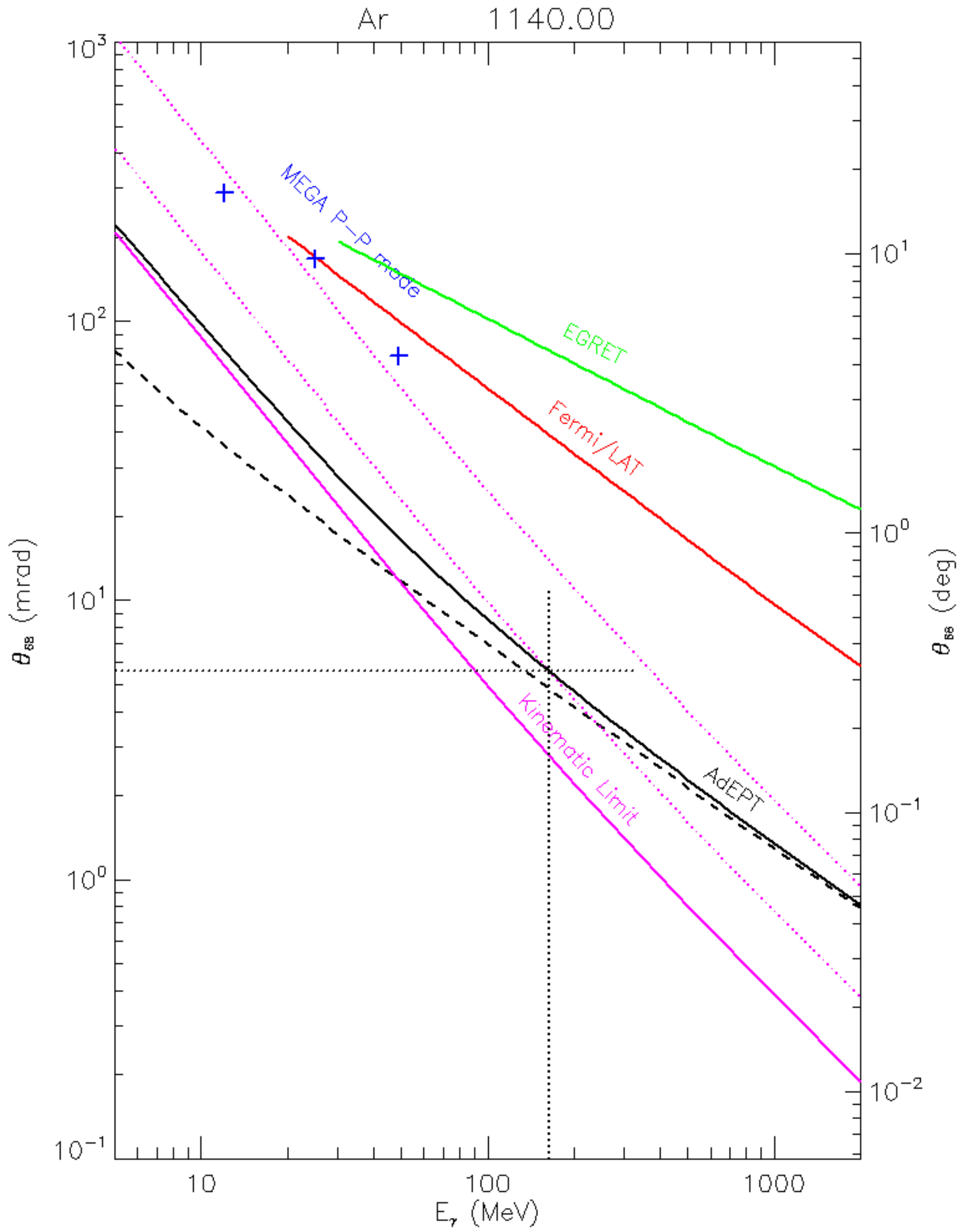

Figure 1 - The calculated angular resolution of the AdEPT telescope concept (solid black line, see §6 and Table 3) as a function of the gamma ray energy is the quadrature sum of the kinematic limit determined for nuclear pair production from [19] (solid magenta line) and the angular resolution limited only by MCS of the electron-positron pair (black dashed line). Twice, and five times the kinematic limit is also shown (dotted magenta lines). Below ~200 MeV, the AdEPT telescope will achieve angular resolution within a factor two of the kinematic limit. The MEGA [29] measured pair production angular resolution (blue crosses), EGRET [73] calibrated angular resolution (green line), and Fermi/LAT front [74] on-orbit angular resolution (red line) are shown for comparison.

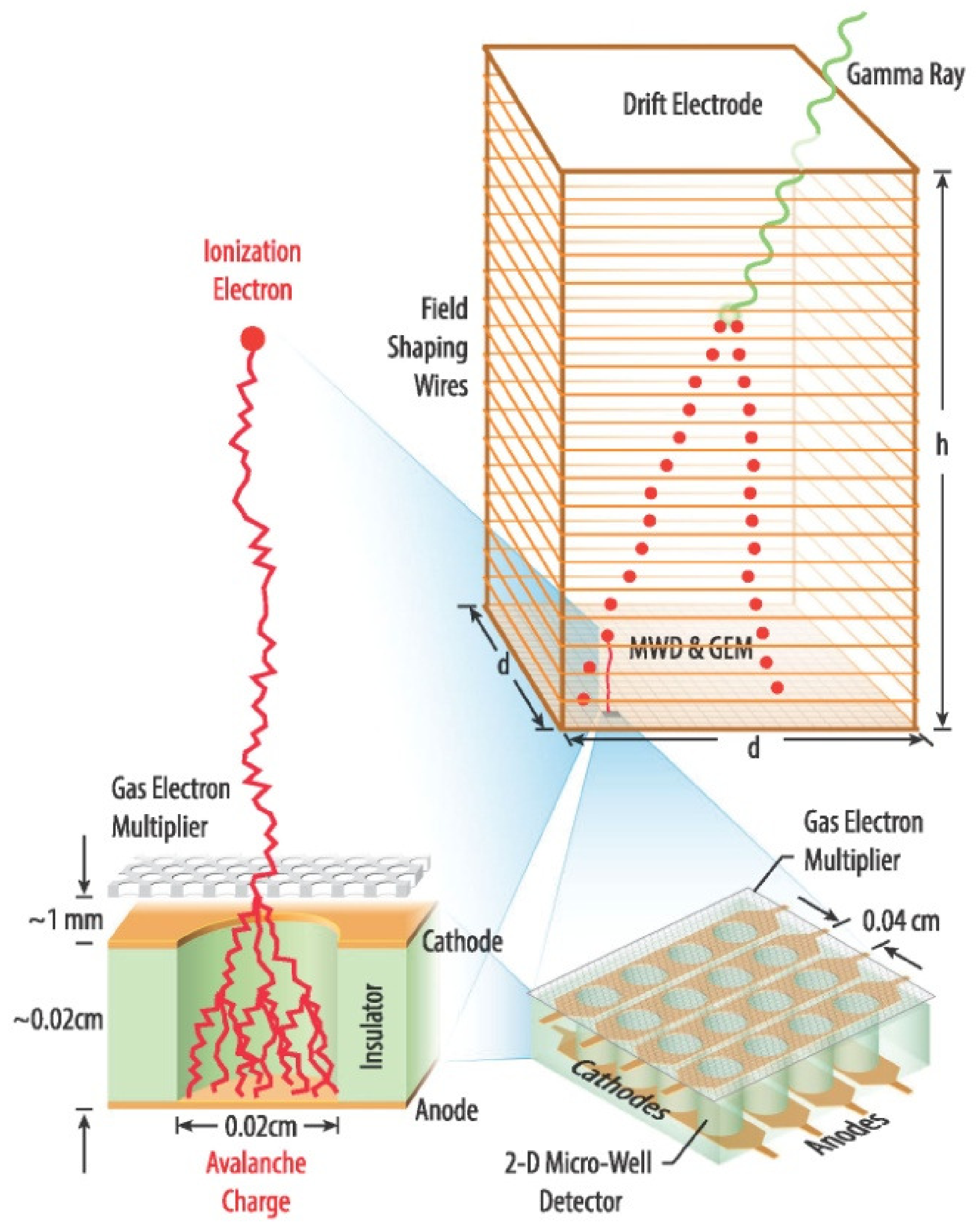

Figure 2 - Schematic of the 3-DTI TPC technology showing how electron-positron pairs are tracked in 3-D. Components of the time projection chamber, micro-well detector plus GEM, and an avalanche in a single well are identitifed.

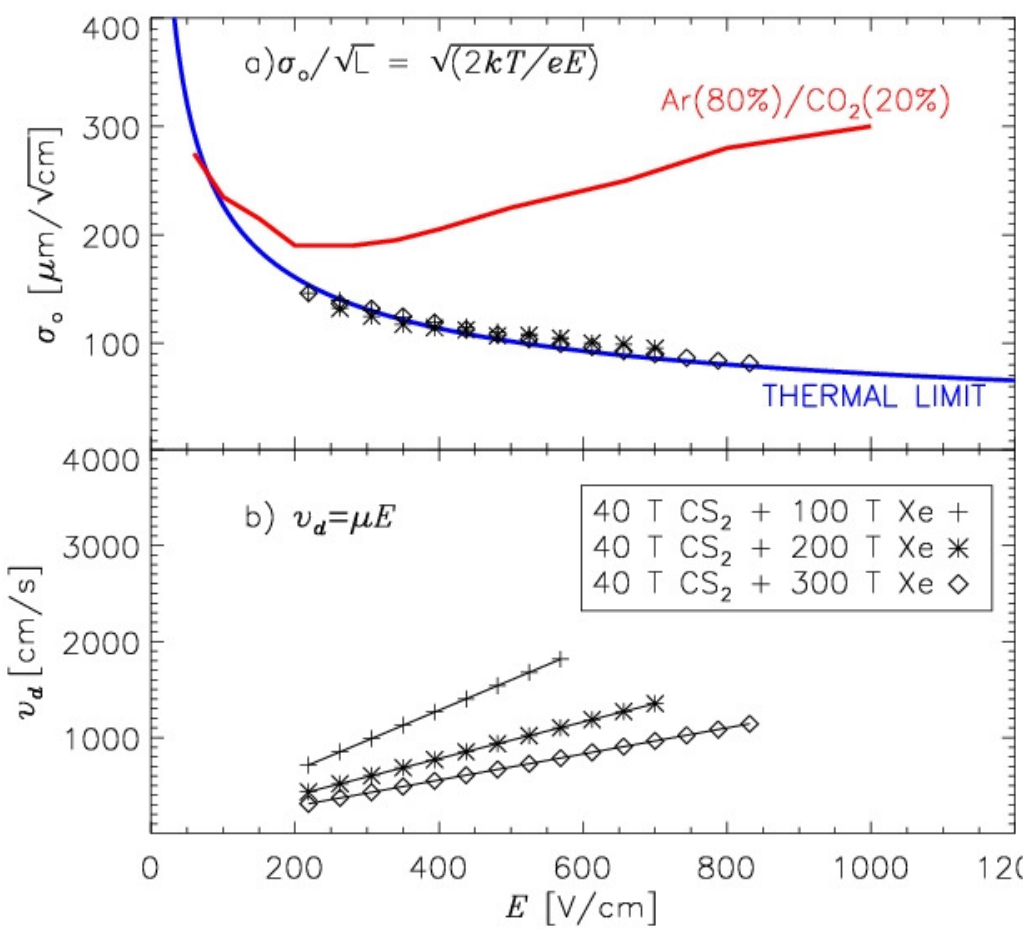

Figure 3 – (top) The longitudinal negative ion diffusion and (bottom) the drift velocity ($v_d = \mu E$) for mixture of Xe + $CS_2$. Thermal diffusion limit for T=300 K is shown as the blue line. Non-thermal diffusion for Ar+$CO_2$ [63] is shown as the red line.

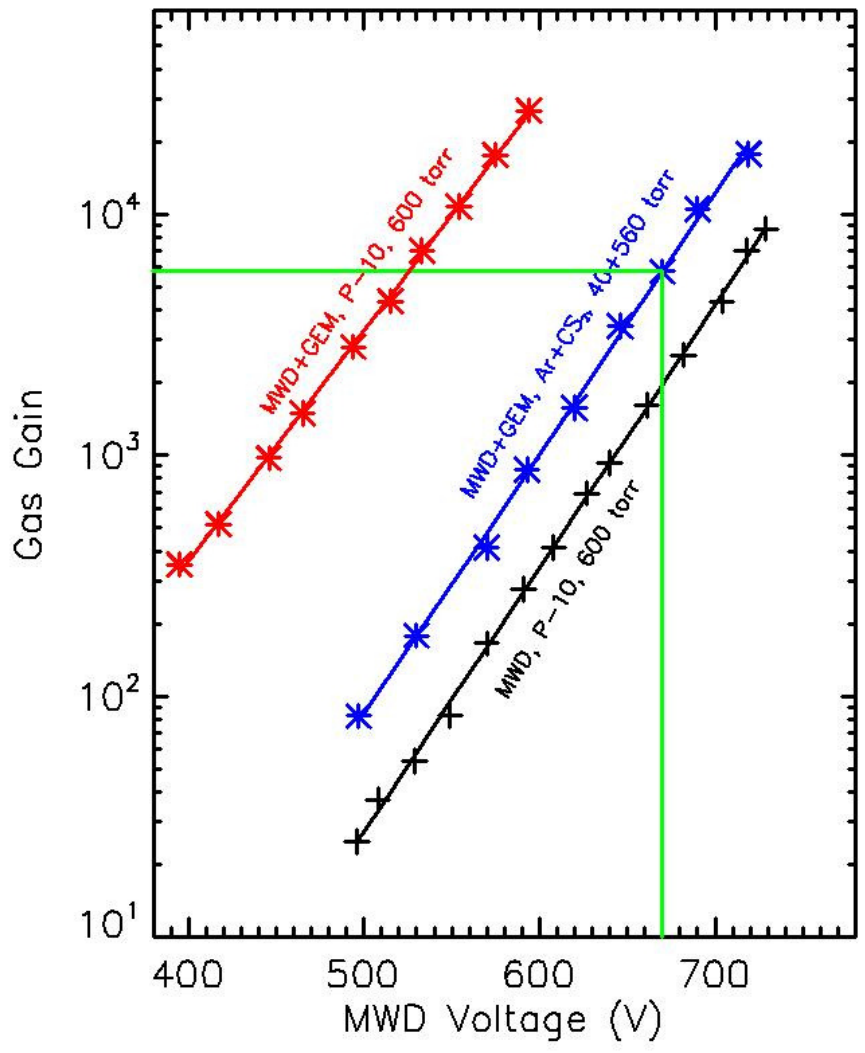

Figure 4 - The gain of our MWD (black data and line) and MWD+GEM (red data and line) as a function of MWD voltage measured in P-10 at 600 torr. The GEM pre-amplifier was operated at a gain of ~100 corresponding to a MWD voltage reduction of ~200 V. The gain of the MWD+GEM as a function of MWD voltage measured in Ar+$CS_2$ (blue data and line) shows that gains in excess of $10^4$ can be achieved. The nominal operating voltage for the MWD and the total gain are indicated by the green lines.

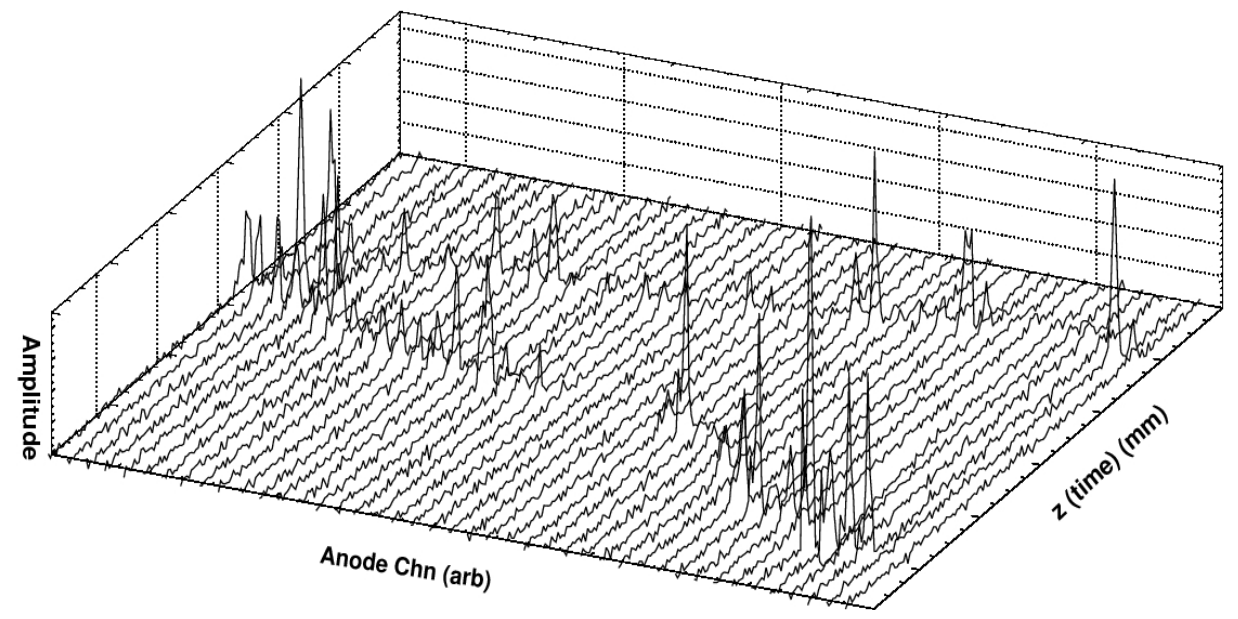

Figure 5 – Electron-positron tracks (X-Z projection) from a 6.129 MeV gamma ray pair interaction in 3-DTI.

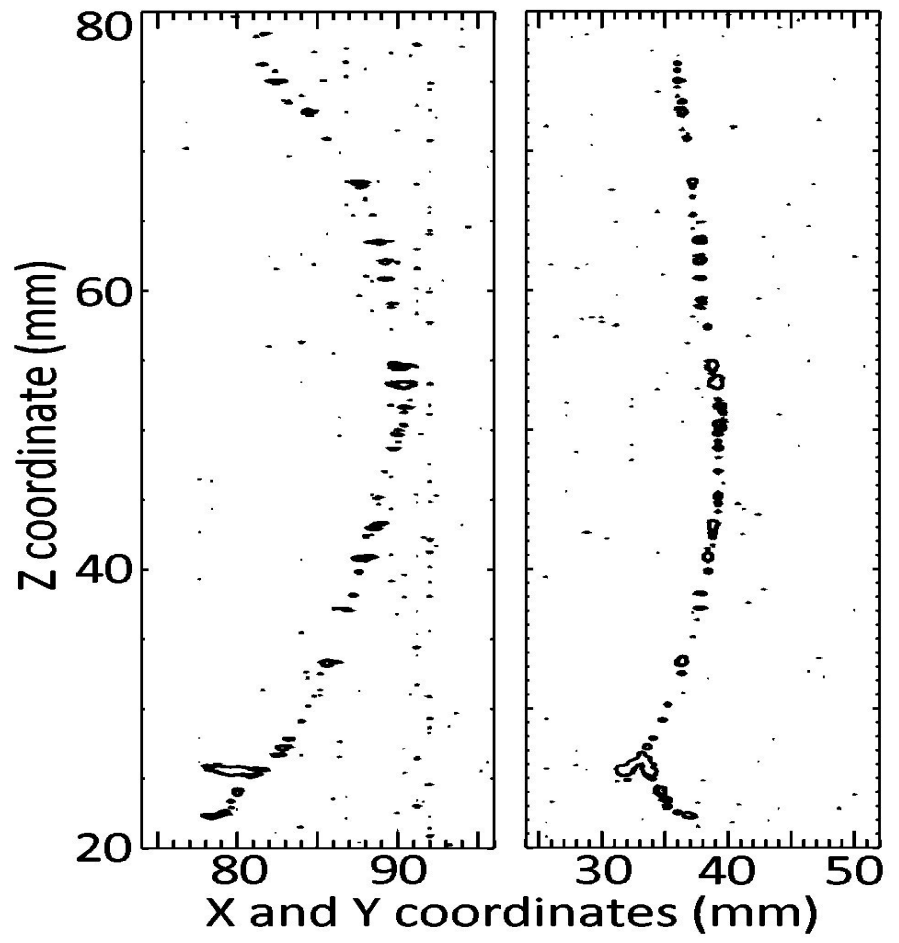

Figure 6 –3-DTI electron track from $^{90}Sr$ in $Ar+CS_2$ (560+40 torr) using a MWD+GEM operating at a total gain of $10^4$. The left and right images are the X-Z and Y-Z projections of the electron track.

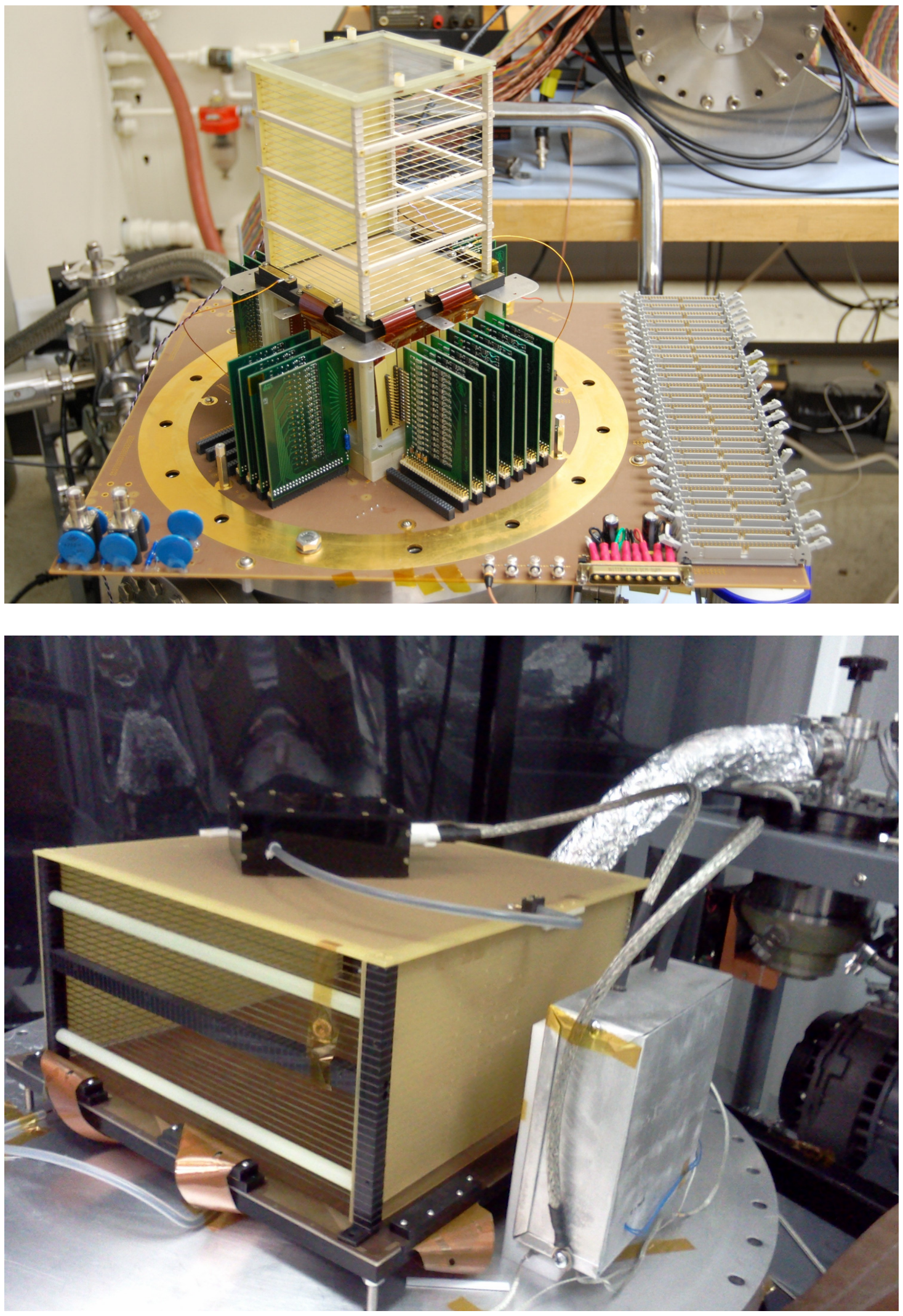

Figure 7 – (top) 10×10×15 $cm^3$ 3-DTI detector consisting of a 10×10 $cm^2$ WMD and 15 cm tall TPC volume. Eleven of the 32 total 16-channel discrete front-end electronics cards are visible. (bottom) 30×30×15 $cm^3$ detector. Three of the 8 flex circuits that connect the MWD to the front-end electronics are visible as well as the drift field high-voltage power (black box) and ripple filter (aluminum box). The upper portions of the vacuum chambers for both detectors were removed for these photos.

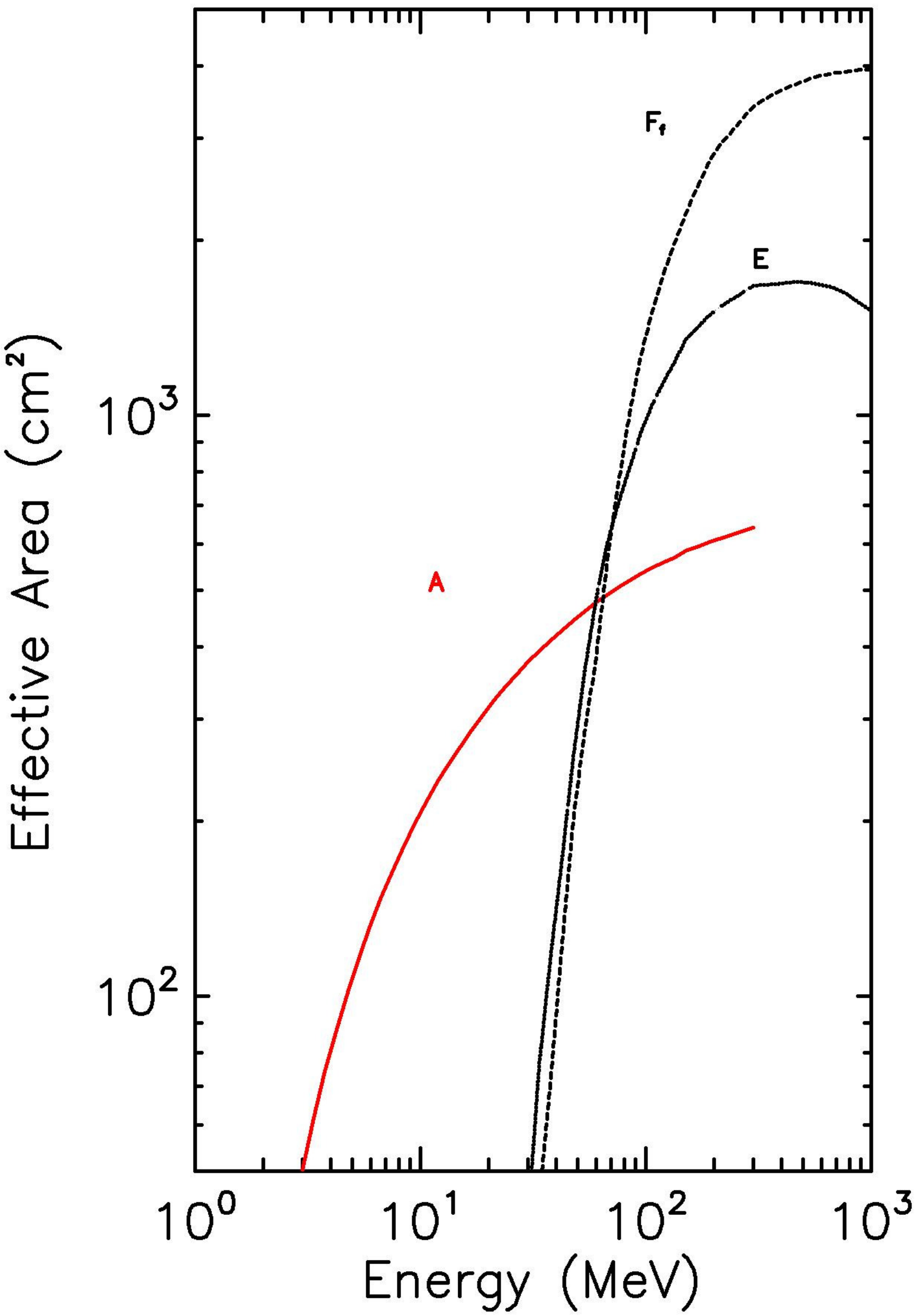

Figure 8 – Effective area of the AdEPT instrument concept (Table 3) (A, solid red lines) compared with EGRET (E, dotted black lines) and FERMI-LAT front detector (Ff, dashed black lines).

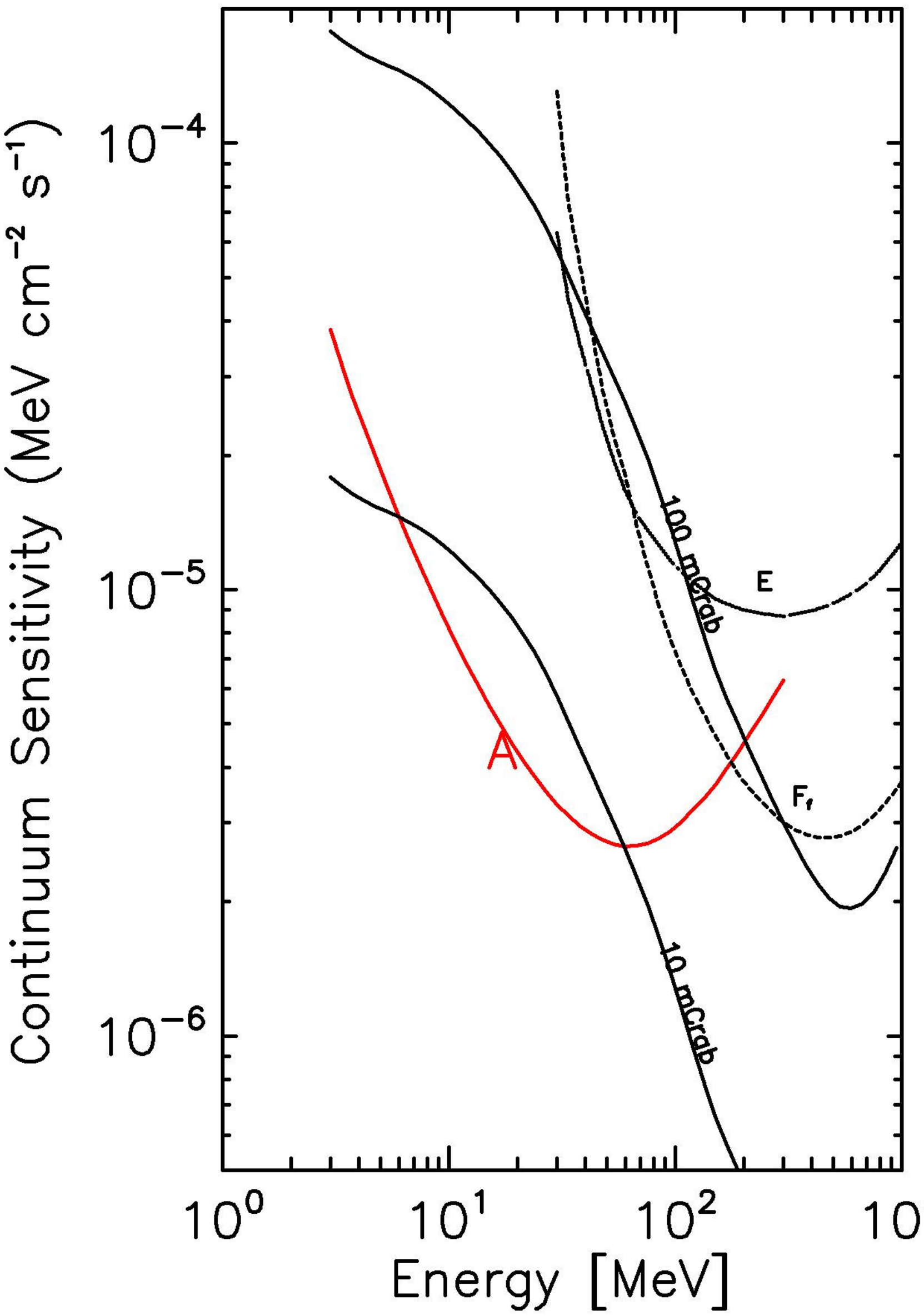

Figure 9 - Continuum sensitivity of the AdEPT instrument concept (Table 3) (A, solid red lines) calculated on the assumption of 3σ significance, $T_{obs}=10^6$ s, ΔE=E, and $F_{egb}=2.7\pm10^{-3}$ (E/1 MeV)$^{-2.1}$. The sensitivity of EGRET (E, dotted black lines), FERMI-LAT front detector ($F_f$, dashed black lines) calculated on the same assumptions. Spectra corresponding to 100 mCrab and 10 mCrab sources are shown for comparison.

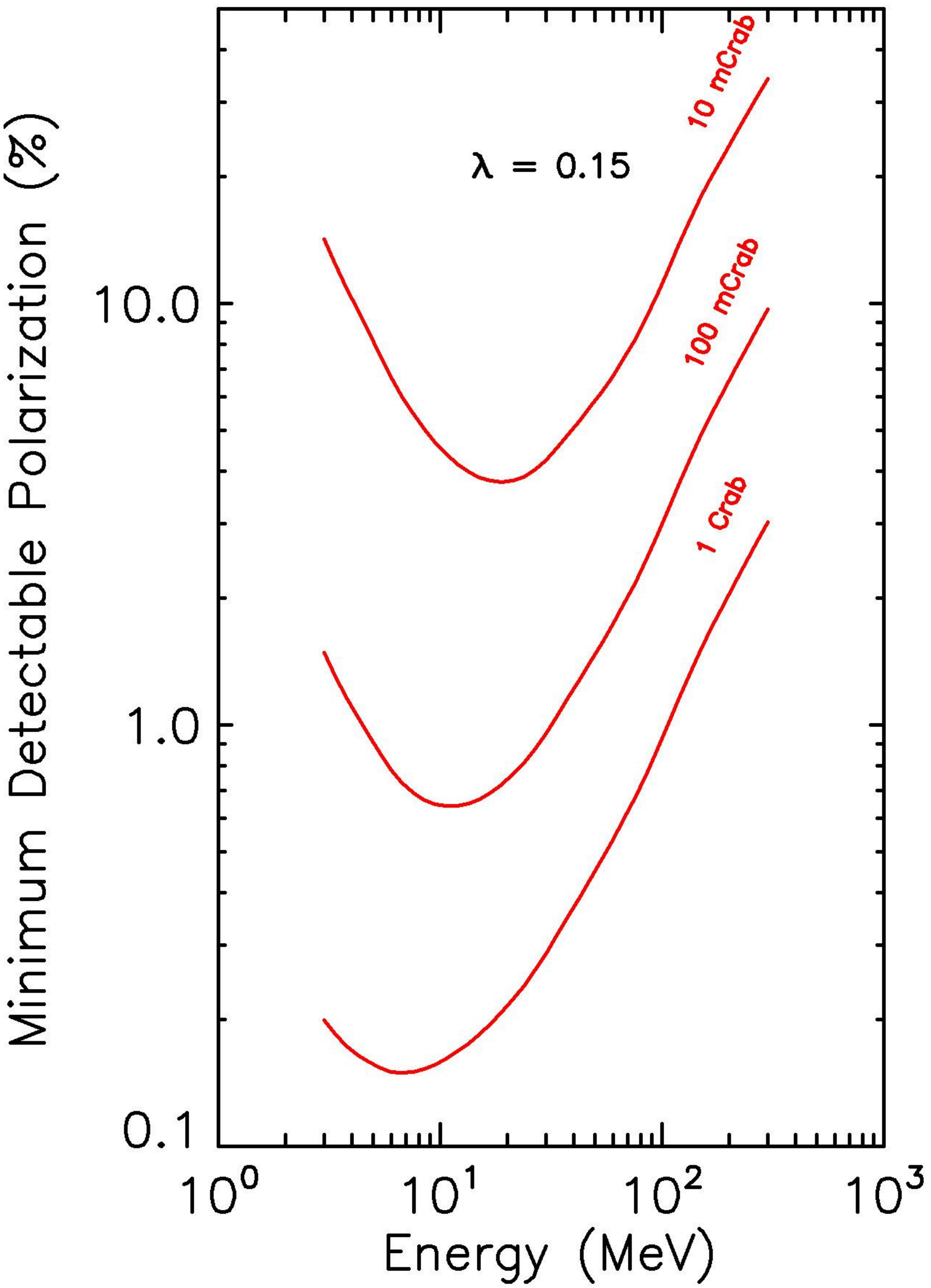

Figure 10 – Minimum detectable polarization of the AdEPT instrument concept (Table 3) calculated on the same assumptions as in Figure 9 for a point source with spectrum and intensity equal to the Crab nebula and for 100 mCrab and 10 mCrab sources.

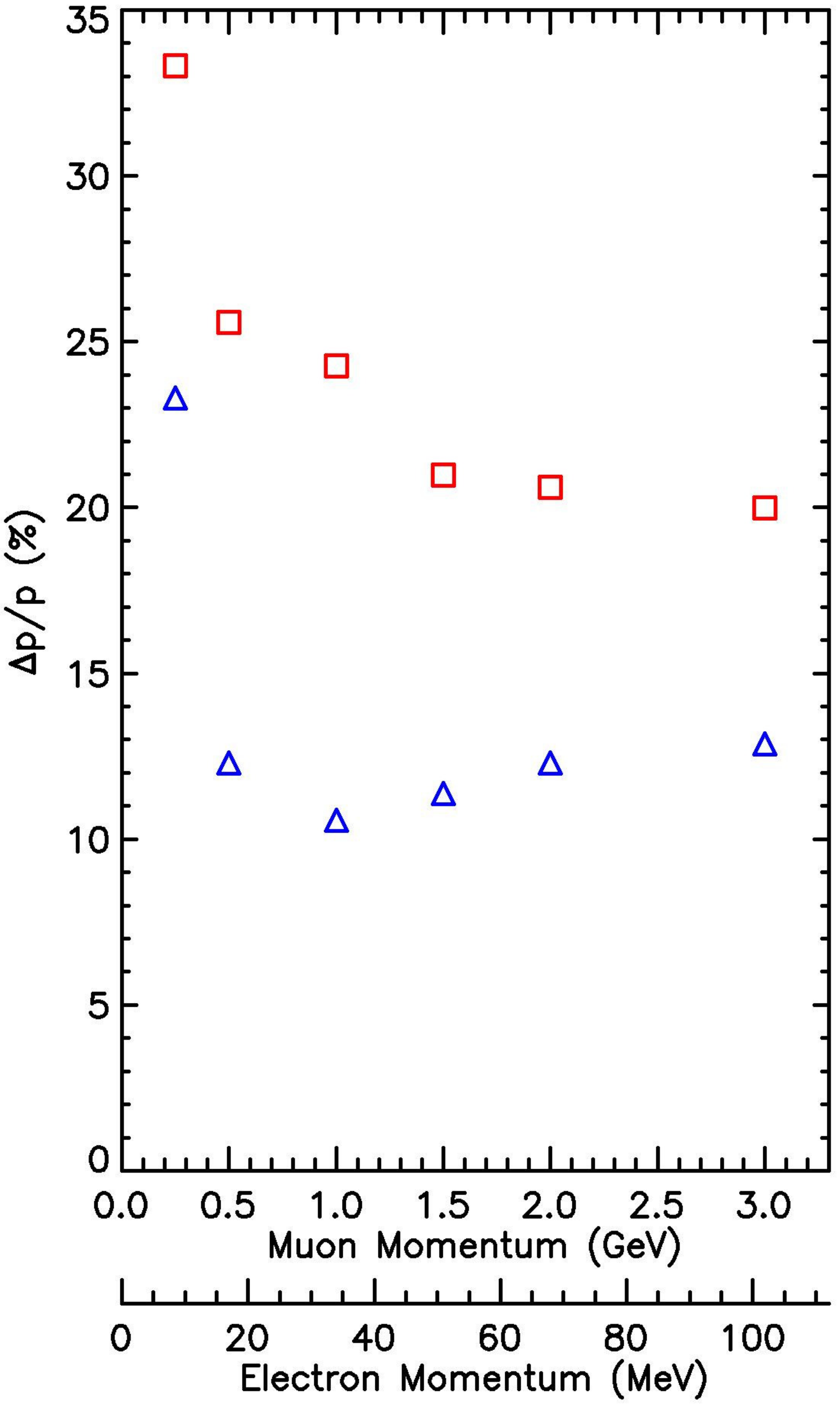

Figure 11 – Monte-Carlo simulation of the ICARUS momentum resolution using classical method (red squares) and the Kalman Filter method (blue triangles). The lower momentum axis has been scaled by the square root of the detector radiation length to give an estimated electron momentum resolution for AdEPT. Adapted from [89].